\documentclass[12pt,a4paper]{article}
\makeatletter
\def\eqnarray{%
\stepcounter{equation}%
\let\@currentlabel=\theequation
\global\@eqnswtrue
\global\@eqcnt\z@
\tabskip\@centering
\let\\=\@eqncr
$$\halign to \displaywidth\bgroup\@eqnsel\hskip\@centering
$\displaystyle\tabskip\z@{##}$&\global\@eqcnt\@ne
\hfil$\displaystyle{{}##{}}$\hfil
&\global\@eqcnt\tw@$\displaystyle\tabskip\z@{##}$\hfil
\tabskip\@centering&\llap{##}\tabskip\z@\cr}
\makeatother
\begin{document}

\title{\sl Universal Yang--Mills Action on Four Dimensional Manifolds}
\author{
  Kazuyuki FUJII
  \thanks{E-mail address : fujii@yokohama-cu.ac.jp }\ \ ,\ \ 
  Hiroshi OIKE
  \thanks{E-mail address : oike@tea.ocn.ne.jp }\quad and\ \ 
  Tatsuo SUZUKI
  \thanks{E-mail address : suzukita@gm.math.waseda.ac.jp }\\
  ${}^{*}$Department of Mathematical Sciences\\
  Yokohama City University\\
  Yokohama, 236--0027\\
  Japan\\
  ${}^{\dagger}$Takado\ 85--5,\ Yamagata, 990--2464\\
  Japan\\
  ${}^{\ddagger}$Department of Mathematical Sciences\\
  Waseda University\\
  Tokyo, 169--8555\\
  Japan\\
  }
\date{}
\maketitle
%
%
%
%
\begin{abstract}
  The usual action of Yang--Mills theory is given by the quadratic form of 
  curvatures of a principal G bundle defined on four dimensional manifolds. 
  The non--linear generalization which is known as the Born--Infeld action 
  has been given. 
  
  In this paper we give another non--linear generalization on four dimensional 
  manifolds and call it a universal Yang--Mills action.
  
  The advantage of our model is that the action splits {\bf automatically} 
  into two parts consisting of self--dual and anti--self--dual directions. 
  Namely, we have automatically the self--dual and anti--self--dual equations 
  without solving the equations of motion as in a usual case. 
  
  Our method may be applicable to recent non--commutative Yang--Mills 
  theories studied widely. 
\end{abstract}
%


%
%
%
%

\section{Introduction}

The purpose of this paper is to develop a universal Yang--Mills theory on 
four dimensional manifolds.

The Yang--Mills theory \cite{YM} is a fundamental one in elementary particle 
physics and there is nothing added furthermore. See also \cite{IM} and 
\cite{WY} as a part of history. We are interested in some non--linear 
generalizations of it. One of them is well--known as the Born--Infeld 
theory \cite{BI}. 

However, we have a dissatisfaction at the model and so we present another 
non--linear generalization. Our action is related to a Chern character for 
a curvature matrix defined in the text, while the Born--Infeld one is related 
to a Chern class for it. Concerning a brief introduction of the Chern theory 
\cite{MN} is recommended.

In the Yang--Mills theory instantons and anti--instantons \cite{BPST} which 
are solutions of the self--dual and anti--self--dual equations play a central 
role. 

The typical characteristic of our model is that the action splits 
{\bf automatically} into two parts consisting of self--dual and 
anti--self--dual directions without solving the equations of motion as 
in a usual case. We hope that our theory may change a framework of 
unified gauge theory.

Although our method in this paper is restricted to four dimensional manifolds 
because of some technical reasons, it must be extended to higher dimensional 
cases (maybe, $4n$--dimensional manifolds). Moreover, it may be applicable to 
non--commutative Yang--Mills theories studied vigorously, see for example 
\cite{AC}, \cite{Nek}. However, it is beyond our scope in the paper. 
Further study will be required.

\section{Mathematical Preliminaries}

We make brief mathematical preliminaries within our necessity 
for the latter convenience. See \cite{MN} for more detailed descriptions. 
In the following we consider the ${\bf C}^{\infty}$ category, namely 
${\bf C}^{\infty}$--manifolds, ${\bf C}^{\infty}$--maps, etc.

Let $M$ be a four dimensional manifold and $G$ a classical group, 
in particular, $U(1)$ and $SU(2)$. We denote by $\{G, P, \pi, M\}$ 
a principal $G$ bundle on $M$
\[
\pi : P\ \longrightarrow\ M,\quad \pi^{-1}(m)\cong G.
\]

We consider a theory of connections of the principal $G$ bundle, so let $g$ be 
the Lie algebra of the group $G$. In the following we treat it locally, 
which is enough for our purpose. 
Let $U$ be any open set in $M$, then a connection $\{A_{\mu}\}$ (a gauge 
potential) is
\[
A_{\mu} : U\ \longrightarrow\ g
\]
and the corresponding curvature $\{F_{\mu\nu}\}$ is given by
\[
F_{\mu\nu}
= \frac{\partial A_{\nu}}{\partial x_{\mu}}-
\frac{\partial A_{\mu}}{\partial x_{\nu}}+[A_{\mu},A_{\nu}]
\equiv \partial_{\mu}A_{\nu}-\partial_{\nu}A_{\mu}+[A_{\mu},A_{\nu}].
\]
We note that 
\[
F_{\mu\nu} : U\ \longrightarrow\ g,\qquad F_{\nu\mu}=-F_{\mu\nu}.
\]

For a map $\phi : U\ \longrightarrow\ G$,\ a gauge transformation by $\phi$ 
is defined by
\[
A_{\mu}\ \longrightarrow\ \phi^{-1}A_{\mu}\phi +\phi^{-1}\partial_{\mu}\phi
\]
and the curvature is then transformed like 
\[
F_{\mu\nu}\ \longrightarrow\ \phi^{-1}F_{\mu\nu}\phi.
\]

\section{Universal Yang--Mills Action : Abelian Case}

In this section we consider the abelian case $G=U(1)$, which gives us a 
prototype. For that let us define a curvature matrix
\begin{equation}
\label{eq:abelian-curvature-matrix}
{\cal F}=
\left(
\begin{array}{cccc}
  0      & F_{12}  & F_{13}  & F_{14} \\
 -F_{12} & 0       & F_{23}  & F_{24} \\
 -F_{13} & -F_{23} & 0       & F_{34} \\
 -F_{14} & -F_{24} & -F_{34} & 0
\end{array}
\right).
\end{equation}

\par \vspace{5mm}
The abelian Yang--Mills action ${\cal A}_{YM}$ (\cite{YM}) \footnote{
it may be suitable to call it the Maxwell action} 
is given by
\begin{equation}
\label{eq:aYM-action}
{\cal A}_{YM}\equiv \frac{1}{2}\mbox{tr} \left(g{\cal F}\right)^{2}
=-g^{2}\sum_{i<j}F_{ij}^{2}
\end{equation}
where $g$ is a coupling constant.

This model is very well--known and there is nothing added furthermore. 
We are interested in some non--linear generalization(s) of it.

\par \vspace{5mm}
A non--linear extension of the abelian Yang--Mills is known as the 
Born--Infeld theory (\cite{BI}) whose action ${\cal A}_{BI}$ is given by
\begin{equation}
\label{eq:BI-action}
 {\cal A}_{BI}\equiv \sqrt{\mbox{det}\left({\bf 1}_{4}+g{\cal F}\right)}
\end{equation}
where $g$ is a coupling constant.

As to this model and its generalizations to non--abelian groups 
there are a lot of papers. However, we don't make a comment in this paper. 

\par \vspace{5mm}
Since we have a dissatisfaction at the model we present another non--linear 
extension. Its action ${\cal A}$ is
\begin{equation}
\label{eq:FOS-action}
{\cal A}_{FOS}\equiv \mbox{tr}\ \mbox{e}^{g{\cal F}}
\end{equation}
where $g$ is a coupling constant \footnote{we can completely determine 
$\mbox{e}^{g{\cal F}}$ itself, see \cite{FOS}}. 
We want to call it a universal Yang--Mills theory

Let us calculate the right hand side of (\ref{eq:FOS-action}). For that 
we must calculate the characteristic polynomial of $g{\cal F}$. 
It is not difficult to show
\begin{eqnarray}
0&=&|\lambda E-g{\cal F}| \nonumber \\
&=&\left|
\begin{array}{cccc}
  \lambda & -gF_{12} & -gF_{13} & -gF_{14} \\
  gF_{12} & \lambda  & -gF_{23} & -gF_{24} \\
  gF_{13} & gF_{23}  & \lambda  & -gF_{34} \\
  gF_{14} & gF_{24}  & gF_{34}  & \lambda
\end{array}
\right|   \nonumber \\
&=&\lambda^{4}+
g^{2}\left(F_{12}^{2}+F_{13}^{2}+F_{14}^{2}+F_{23}^{2}+F_{24}^{2}+
F_{34}^{2}\right)\lambda^{2}+
g^{4}\left(F_{12}F_{34}-F_{13}F_{24}+F_{14}F_{23}\right)^{2}.
\end{eqnarray}
We set $t=\lambda^{2}$ for simplicity. Then
\[
t^{2}+g^{2}\left(F_{12}^{2}+F_{13}^{2}+F_{14}^{2}+F_{23}^{2}+F_{24}^{2}+
F_{34}^{2}\right)t+
g^{4}\left(F_{12}F_{34}-F_{13}F_{24}+F_{14}F_{23}\right)^{2}=0.
\]

Next let us calculate the discriminant.
\begin{eqnarray*}
D
&=&g^{4}\left(F_{12}^{2}+F_{13}^{2}+F_{14}^{2}+F_{23}^{2}+F_{24}^{2}+
F_{34}^{2}\right)^{2}-
4g^{4}\left(F_{12}F_{34}-F_{13}F_{24}+F_{14}F_{23}\right)^{2} \\
&=&g^{4}
\left\{F_{12}^{2}+F_{13}^{2}+F_{14}^{2}+F_{23}^{2}+F_{24}^{2}+F_{34}^{2}
-2(F_{12}F_{34}-F_{13}F_{24}+F_{14}F_{23})\right\} \times \\
&&\ \ \
\left\{F_{12}^{2}+F_{13}^{2}+F_{14}^{2}+F_{23}^{2}+F_{24}^{2}+F_{34}^{2}
+2(F_{12}F_{34}-F_{13}F_{24}+F_{14}F_{23})\right\} \\
&=&g^{4}
\left\{(F_{12}-F_{34})^{2}+(F_{13}+F_{24})^{2}+(F_{14}-F_{23})^{2}\right\}
\times \\
&&\ \ \
\left\{(F_{12}+F_{34})^{2}+(F_{13}-F_{24})^{2}+(F_{14}+F_{23})^{2}\right\}.
\end{eqnarray*}

\par \noindent
If we set
\begin{eqnarray}
X_{sd}^{2}&=& 
(F_{12}-F_{34})^{2}+(F_{13}+F_{24})^{2}+(F_{14}-F_{23})^{2}, \\
X_{asd}^{2}&=& 
(F_{12}+F_{34})^{2}+(F_{13}-F_{24})^{2}+(F_{14}+F_{23})^{2}
\end{eqnarray}
then 
\[
D=g^{4}X_{sd}^{2}X_{asd}^{2}.
\]
It is easy to see
\begin{eqnarray*}
X_{asd}^{2}+X_{sd}^{2}
&=&2\left(F_{12}^{2}+F_{13}^{2}+F_{14}^{2}+F_{23}^{2}+F_{24}^{2}+
F_{34}^{2}\right), \\
X_{asd}^{2}-X_{sd}^{2}
&=&4\left(F_{12}F_{34}-F_{13}F_{24}+F_{14}F_{23}\right),
\end{eqnarray*}
so 
\[
t^{2}+g^{2}\frac{X_{sd}^{2}+X_{asd}^{2}}{2}t+
g^{4}\frac{(X_{asd}^{2}-X_{asd}^{2})^{2}}{16}=0.
\]

\par \vspace{3mm} \noindent
From this we have
\begin{eqnarray*}
t_{\pm}
&=&g^{2}\left\{-\frac{X_{sd}^{2}+X_{asd}^{2}}{4}
\pm \frac{X_{sd}X_{asd}}{2}\right\}
=\frac{g^{2}}{4}\left\{-(X_{sd}^{2}+X_{asd}^{2})\pm 2X_{sd}X_{asd}\right\} \\
&=&-\frac{g^{2}}{4}\left\{(X_{sd}^{2}+X_{asd}^{2})\mp 2X_{sd}X_{asd}\right\}
=-\frac{g^{2}}{4}\left(X_{sd}\mp X_{asd}\right)^{2},
\end{eqnarray*}
therefore 
\begin{eqnarray}
\lambda_{1}&=&i\frac{g}{2}(X_{sd}+X_{asd}),\quad 
\lambda_{2}=-i\frac{g}{2}(X_{sd}+X_{asd}),     \nonumber \\
\lambda_{3}&=&i\frac{g}{2}(X_{sd}-X_{asd}),\quad
\lambda_{4}=-i\frac{g}{2}(X_{sd}-X_{asd})
\end{eqnarray}
because of $\lambda^{2}=t$.

\par \vspace{5mm} \noindent
As a result we obtain
\begin{eqnarray}
\label{eq:abelian-result}
{\cal A}_{FOS}
&\equiv& \mbox{tr}\ \mbox{e}^{g{\cal F}} \nonumber \\
&=&\mbox{e}^{\lambda_{1}}+\mbox{e}^{\lambda_{2}}+\mbox{e}^{\lambda_{4}}+
\mbox{e}^{\lambda_{4}}  \nonumber \\
&=&2\cos\left(g\frac{X_{sd}+X_{asd}}{2}\right)+
   2\cos\left(g\frac{X_{sd}-X_{asd}}{2}\right)   \nonumber \\
&=&4\cos\left(g\frac{X_{sd}}{2}\right)\cos\left(g\frac{X_{asd}}{2}\right).
\end{eqnarray}
It is very notable that
\begin{eqnarray}
&&X_{sd}=0 \Longleftrightarrow 
F_{12}=F_{34},\ F_{13}=-F_{24},\ F_{14}=F_{23} \Longleftrightarrow 
\{F_{ij}\}\ \mbox{is {\bf self--dual}}, \\
&&X_{asd}=0 \Longleftrightarrow 
F_{12}=-F_{34},\ F_{13}=F_{24},\ F_{14}=-F_{23} \Longleftrightarrow 
\{F_{ij}\}\ \mbox{is {\bf anti--self--dual}}.
\end{eqnarray}

The characteristic of our model is that the action (\ref{eq:abelian-result}) 
splits {\bf automatically} into two parts consisting of self--dual and 
anti--self--dual directions. 
Namely, we have {\bf automatically} the self--dual and anti--self--dual 
equations without solving the equations of motion as in a usual case.

\par \vspace{3mm}
Last in this section let us make an important comment on the action 
(\ref{eq:abelian-result}). Since the target (as a map) of curvatures 
$\{F_{ij}\}$ is the Lie algebra $u(1)=\sqrt{-1}{\bf R}$ it may be 
appropriate to write $F_{ij}=\sqrt{-1}G_{ij}$. 
Then (\ref{eq:abelian-result}) is changed into the more suitable form
\begin{equation}
\label{eq:abelian-result-modify}
{\cal A}_{FOS}=
4\cosh\left(g\frac{Y_{sd}}{2}\right)\cosh\left(g\frac{Y_{asd}}{2}\right)
\end{equation}
with
\begin{eqnarray}
Y_{sd}^{2}&=& 
(G_{12}-G_{34})^{2}+(G_{13}+G_{24})^{2}+(G_{14}-G_{23})^{2}\ \geq\ 0, \\
Y_{asd}^{2}&=& 
(G_{12}+G_{34})^{2}+(G_{13}-G_{24})^{2}+(G_{14}+G_{23})^{2}\ \geq\ 0.
\end{eqnarray}

The same thing will hold for {\bf non--abelian cases} in the following.

\section{Universal Yang--Mills Action : Non--Abelian Case}

In this section we consider the non--abelian case $G=SU(2)$. From the form in 
(\ref{eq:abelian-result}) it is natural to conjecture
\begin{equation}
\label{eq:non-abelian-result}
{\cal A}_{FOS}= 4\ \mbox{tr}
\cos\left(g\frac{X_{sd}}{2}\right)\cos\left(g\frac{X_{asd}}{2}\right)
\end{equation}
where
\begin{eqnarray}
X_{sd}^{2}&=& 
(F_{12}-F_{34})^{2}+(F_{13}+F_{24})^{2}+(F_{14}-F_{23})^{2}, \\
X_{asd}^{2}&=& 
(F_{12}+F_{34})^{2}+(F_{13}-F_{24})^{2}+(F_{14}+F_{23})^{2}.
\end{eqnarray}
However, a natural question arises. What is a curvature matrix giving 
(\ref{eq:non-abelian-result}) as a result ?  This is not an easy problem.

As a first trial we generalize (\ref{eq:abelian-curvature-matrix}) as
\begin{equation}
\label{eq:non-abelian-curvature-matrix}
{\cal F}=
\left(
\begin{array}{cccc}
 {\bf 0}_{2} & F_{12} & F_{13} & F_{14}    \\
 -F_{12} & {\bf 0}_{2} & F_{23} & F_{24}   \\
 -F_{13} & -F_{23} & {\bf 0}_{2} & F_{34}  \\
 -F_{14} & -F_{24} & -F_{34} & {\bf 0}_{2} 
\end{array}
\right)
\end{equation}
where 
\begin{equation}
F_{ij}=F_{ij}^{1}\sigma_{1}+F_{ij}^{2}\sigma_{2}+F_{ij}^{3}\sigma_{3}
\ :\ U \subset M \longrightarrow\ su(2)
\end{equation}
with the Pauli matrices $\{\sigma_{1}, \sigma_{2}, \sigma_{3}\}$ defined by
\begin{equation}
\sigma_{1} = 
\left(
  \begin{array}{cc}
    0 & 1 \\
    1 & 0
  \end{array}
\right), \quad 
\sigma_{2} = 
\left(
  \begin{array}{cc}
    0 & -i \\
    i & 0
  \end{array}
\right), \quad 
\sigma_{3} = 
\left(
  \begin{array}{cc}
    1 & 0 \\
    0 & -1
  \end{array}
\right).
\end{equation}

In the following it is better to use
\begin{equation}
F_{ij}\ \longrightarrow\ {\widehat{F}}_{ij}=
\left(
  \begin{array}{c}
    F_{ij}^{1} \\
    F_{ij}^{2} \\
    F_{ij}^{3}
  \end{array}
\right),
\end{equation}
so we set for simplicity 
\begin{equation}
{\bf a}\equiv {\widehat{F}}_{12},\ \
{\bf b}\equiv {\widehat{F}}_{13},\ \
{\bf c}\equiv {\widehat{F}}_{14},\ \
{\bf d}\equiv {\widehat{F}}_{23},\ \
{\bf e}\equiv {\widehat{F}}_{24},\ \
{\bf f}\equiv {\widehat{F}}_{34}.
\end{equation}

Here we prepare the notations : for three three--dimensional vectors 
${\bf x}$,\ ${\bf y}$,\ ${\bf z}$ we set the inner products, 
exterior products and scalar triple product respectively as
\begin{eqnarray*}
&&
({\bf x}{\bf y})\equiv {\bf x}\cdot {\bf y},\quad 
({\bf x}{\bf x})\equiv {\bf x}^{2},\quad 
({\bf y}{\bf y})\equiv {\bf y}^{2}, \\
&&[{\bf x}{\bf y}]\equiv {\bf x}\wedge {\bf y},\quad \mbox{etc},\\
&&
|{\bf x}{\bf y}{\bf z}|\equiv ({\bf x}\ [{\bf y}{\bf z}])=
{\bf x}\cdot {\bf y}\wedge {\bf z}=|{\bf x},\ {\bf y},\ {\bf z}|.
\end{eqnarray*}
Now we write down a useful relation : for
\[
A=a^{1}\sigma_{1}+a^{2}\sigma_{2}+a^{3}\sigma_{3}
\ \Longrightarrow\ {\bf a}=(a^{1},a^{2},a^{3})^{T}
\]
we have
\begin{eqnarray}
\label{eq:important-relation}
A^{2}
&=&\left\{(a^{1})^{2}+(a^{2})^{2}+(a^{3})^{2}\right\}{\bf 1}_{2}
+a^{1}a^{2}\{\sigma_{1},\sigma_{2}\}
+a^{1}a^{3}\{\sigma_{1},\sigma_{3}\}
+a^{2}a^{3}\{\sigma_{2},\sigma_{3}\} \nonumber \\
&=&\left\{(a^{1})^{2}+(a^{2})^{2}+(a^{3})^{2}\right\}{\bf 1}_{2}
={\bf a}^{2}{\bf 1}_{2}.
\end{eqnarray}

\par \vspace{5mm}
By way of trial we define the universal Yang--Mills action ${\cal A}$ like
\begin{equation}
\label{eq:non-abelian-action}
{\cal A}_{FOS}\equiv \mbox{Tr}\ \mbox{e}^{g{\cal F}}
\end{equation}
with (\ref{eq:non-abelian-curvature-matrix}), where $g$ is a coupling 
constant. Here, to avoid a confusion we used the notation Tr not tr in 
(\ref{eq:FOS-action}) because each component of ${\cal F}$ in 
(\ref{eq:non-abelian-curvature-matrix}) is a matrix.

\par \vspace{3mm}
Let us calculate the right hand side of (\ref{eq:non-abelian-action}), which 
is hard task this time. 
The characteristic polynomial of $g{\cal F}$ is 
\begin{eqnarray}
\label{eq:characteristic-polynomial-su(2)}
0&=&|\lambda E-g{\cal F}| \nonumber \\
&=&\left|
\begin{array}{cccc}
  \lambda{\bf 1}_{2} & -gF_{12} & -gF_{13} & -gF_{14} \\
  gF_{12} & \lambda{\bf 1}_{2}  & -gF_{23} & -gF_{24} \\
  gF_{13} & gF_{23}  & \lambda{\bf 1}_{2}  & -gF_{34} \\
  gF_{14} & gF_{24}  & gF_{34}  & \lambda{\bf 1}_{2}
\end{array}
\right|   \nonumber \\
&=&
\left[
\lambda^{4}+
g^{2}\left({\bf a}^{2}+{\bf b}^{2}+{\bf c}^{2}+{\bf d}^{2}+{\bf e}^{2}+
{\bf f}^{2}\right)\lambda^{2}-
2ig^{3}\left(|{\bf a}{\bf b}{\bf d}|+|{\bf a}{\bf c}{\bf e}|+
|{\bf b}{\bf c}{\bf f}|+|{\bf d}{\bf e}{\bf f}|\right)\lambda +
\right. \nonumber \\
&&{}\left. g^{4}\mbox{PF}({\cal F})\right]^{2}
\end{eqnarray}
where $\mbox{PF}({\cal F})$ is the Pfaffian of ${\cal F}$ and is given by 
\begin{eqnarray}
\mbox{PF}({\cal F})
&=&{\bf a}^{2}{\bf f}^{2}+{\bf b}^{2}{\bf e}^{2}+{\bf c}^{2}{\bf d}^{2}-
2\left\{
({\bf a}{\bf b})({\bf e}{\bf f})-({\bf a}{\bf f})({\bf b}{\bf e})+
({\bf a}{\bf e})({\bf b}{\bf f})-({\bf a}{\bf c})({\bf d}{\bf f})+
\right. \nonumber \\
&&\left.
({\bf a}{\bf f})({\bf c}{\bf d})-({\bf a}{\bf d})({\bf c}{\bf f})+
({\bf b}{\bf c})({\bf d}{\bf e})-({\bf b}{\bf e})({\bf c}{\bf d})+
({\bf b}{\bf d})({\bf c}{\bf e})
\right\}.
\end{eqnarray}
In the calculation the relation (\ref{eq:important-relation}) plays an 
important role. 

\par \vspace{3mm}
Here by neglecting the $\lambda$--term \footnote{we can formally solve 
(\ref{eq:characteristic-polynomial-su(2)}) by use of the formula given by 
Ferrari or Euler, see for example \cite{KF3}} we have
\[
t^{2}+
g^{2}\left({\bf a}^{2}+{\bf b}^{2}+{\bf c}^{2}+{\bf d}^{2}+{\bf e}^{2}+
{\bf f}^{2}\right)t+
g^{4}\mbox{PF}({\cal F})=0
\]
for $t=\lambda^{2}$. Let us calculate the discriminant
\begin{eqnarray*}
D
&=&g^{4}\left({\bf a}^{2}+{\bf b}^{2}+{\bf c}^{2}+{\bf d}^{2}+{\bf e}^{2}+
{\bf f}^{2}\right)^{2}-
4g^{4}\mbox{PF}({\cal F})^{2} \\
&=&g^{4}
\left\{{\bf a}^{2}+{\bf b}^{2}+{\bf c}^{2}+{\bf d}^{2}+{\bf e}^{2}+
{\bf f}^{2}+2\mbox{PF}({\cal F})\right\} \times \\
&&\ \ \
\left\{{\bf a}^{2}+{\bf b}^{2}+{\bf c}^{2}+{\bf d}^{2}+{\bf e}^{2}+
{\bf f}^{2}-2\mbox{PF}({\cal F})\right\}. \\
\end{eqnarray*}

\par \noindent
Here we define 
\begin{eqnarray}
\tilde{X}_{sd}^{2}&=&
\left\{
({\bf a}-{\bf f})^{2}+({\bf b}+{\bf e})^{2}+({\bf c}-{\bf d})^{2}
\right\}, \\
\tilde{X}_{asd}^{2}&=&
\left\{
({\bf a}+{\bf f})^{2}+({\bf b}-{\bf e})^{2}+({\bf c}+{\bf d})^{2}
\right\}
\end{eqnarray}
for the latter convenience. For
\[
\tilde{X}_{sd}^{2}\tilde{X}_{asd}^{2}=
\left\{
({\bf a}-{\bf f})^{2}+({\bf b}+{\bf e})^{2}+({\bf c}-{\bf d})^{2}
\right\}
\left\{
({\bf a}+{\bf f})^{2}+({\bf b}-{\bf e})^{2}+({\bf c}+{\bf d})^{2}
\right\}
\]
a long calculation leads to
\begin{eqnarray}
D-g^{4}\tilde{X}_{sd}^{2}\tilde{X}_{asd}^{2}=
&&-4g^{4}
\left[
\left([{\bf a}{\bf f}]-[{\bf b}{\bf e}]+[{\bf c}{\bf d}]\right)^{2}
\right. \nonumber \\
&&\left.
\qquad -4\left\{([{\bf a}{\bf b}][{\bf e}{\bf f}])-
([{\bf a}{\bf c}][{\bf d}{\bf f}])+([{\bf b}{\bf c}][{\bf d}{\bf e}])
\right\}
\right].
\end{eqnarray}
Therefore we have
\begin{equation}
D=g^{4}\tilde{X}_{sd}^{2}\tilde{X}_{asd}^{2}\quad 
\mbox{mod\ \{Exterior Products\}}.
\end{equation}

\par \vspace{3mm}
Since 
\[
|{\bf a}{\bf b}{\bf d}|+|{\bf a}{\bf c}{\bf e}|+
|{\bf b}{\bf c}{\bf f}|+|{\bf d}{\bf e}{\bf f}|
=0\quad \mbox{mod\ \{Exterior Products\}},
\]
the equation
\begin{equation}
0=|\lambda E-g{\cal F}| \quad \mbox{mod\ \{Exterior Products\}}
\end{equation}
gives
\[
t^{2}+g^{2}\frac{\tilde{X}_{sd}^{2}+\tilde{X}_{asd}^{2}}{2}t+
g^{4}\frac{(\tilde{X}_{asd}^{2}-\tilde{X}_{asd}^{2})^{2}}{16}=0
\]
and so
\[
t_{\pm}=-\frac{g^{2}}{4}\left(\tilde{X}_{sd}\mp \tilde{X}_{asd}\right)^{2}.
\]
Therefore
\begin{eqnarray}
\lambda_{1}&=&i\frac{g}{2}(\tilde{X}_{sd}+\tilde{X}_{asd}),\quad 
\lambda_{2}=-i\frac{g}{2}(\tilde{X}_{sd}+\tilde{X}_{asd}),  \nonumber \\
\lambda_{3}&=&i\frac{g}{2}(\tilde{X}_{sd}-\tilde{X}_{asd}),\quad
\lambda_{4}=-i\frac{g}{2}(\tilde{X}_{sd}-\tilde{X}_{asd}).
\end{eqnarray}
That is, the solutions are 
$\{\lambda_{1},\lambda_{1},\lambda_{2},\lambda_{2},\lambda_{3},\lambda_{3},
\lambda_{4},\lambda_{4}\}$ because they are two--fold.

\par \vspace{3mm} \noindent
Finally we obtain the desired result
\begin{eqnarray}
\label{eq:non-abelian-calculation}
{\cal A}_{FOS}&\equiv& \mbox{Tr}\ \mbox{e}^{g{\cal F}}\quad 
\mbox{mod\ \{Exterior Products\}} \nonumber \\
&=&2\left(
\mbox{e}^{\lambda_{1}}+\mbox{e}^{\lambda_{2}}+\mbox{e}^{\lambda_{4}}+
\mbox{e}^{\lambda_{4}}\right)  \nonumber \\
&=&8\cos\left(g\frac{\tilde{X}_{sd}}{2}\right)
    \cos\left(g\frac{\tilde{X}_{asd}}{2}\right)
\nonumber \\
&=&4\ \mbox{tr}\cos\left(g\frac{X_{sd}}{2}\right)
             \cos\left(g\frac{X_{asd}}{2}\right).
\end{eqnarray}

\par \vspace{5mm}
A comment is in order.\ We would like to point out a similarity between our 
idea and a topic in Topology. 
The first homology group $H_{1}(M,{\bf Z})$ of $M$ is an abelian group, 
while the first homotopy group (fundamental group) 
$\pi_{1}(M)$ of $M$ is in general non--abelian. However, there is a famous 
relation 
\[
H_{1}(M,{\bf Z})\cong \pi_{1}(M)/[\pi_{1}(M),\pi_{1}(M)]
=\pi_{1}(M)\ \mbox{mod}\ [\pi_{1}(M),\pi_{1}(M)]
\]
where $[\pi_{1}(M),\pi_{1}(M)]$ is the commutator subgroup generated by 
$\{aba^{-1}b^{-1}\ |\ a,b \in \pi_{1}(M) \}$. See for example \cite{MN}.\quad
Our method is {\bf a kind of abelianization}.

\par \vspace{3mm}
It should be noted that the gauge groups of our model are restricted to 
$U(1)$ and $SU(2)$, which is of course not enough and we must treat the 
general $SU(n)$ for $n \in {\bf N}$. It is natural to conjecture 
\[
{\cal A}_{FOS}= 4\ \mbox{tr}
\cos\left(g\frac{X_{sd}}{2}\right)\cos\left(g\frac{X_{asd}}{2}\right)
\]
with
\begin{eqnarray*}
F_{ij} &:& U \subset M\ \longrightarrow\ su(n), \\
X_{sd}^{2}&=&
(F_{12}-F_{34})^{2}+(F_{13}+F_{24})^{2}+(F_{14}-F_{23})^{2}, \\
X_{asd}^{2}&=& 
(F_{12}+F_{34})^{2}+(F_{13}-F_{24})^{2}+(F_{14}+F_{23})^{2}.
\end{eqnarray*}

At the present time we don't know how to derive this action from a curvature 
matrix like (\ref{eq:non-abelian-curvature-matrix}). We leave it to some 
readers as a challenging problem.

\par \vspace{3mm}
Our model is also based on four dimensional manifolds $M$, which is too 
restrictive in the view point of M--theory or F--theory. We want to generalize 
$M$ to a $4n$--dimensional manifold and treat an action like 
(\ref{eq:non-abelian-calculation}). 

Moreover, it may be interesting to apply our method to non--commutative 
Yang--Mills theory (for example \cite{AC}, \cite{Nek}), which is however 
beyond our scope in this paper.

\section{Discussion}

In this letter we presented the non--linear generalization of the 
Yang--Mills theory which is different from the Born--Infeld theory. Our 
action is related to a Chern Character, while that of Born--Infeld is 
related to a Chern class. Our model has an advantage. 

\par \noindent 
Details of the paper and further developments including some quantization 
will be published elsewhere \cite{FOS}.

\par \vspace{3mm}
Last let us make a brief comment. We are studying a quantum computation based 
on Cavity QED whose image is

\vspace{10mm}
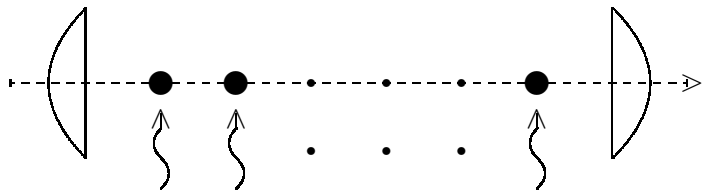
\begin{center}
\setlength{\unitlength}{1mm} 
\begin{picture}(110,40)(0,-20)
\bezier{200}(20,0)(10,10)(20,20)
\put(20,0){\line(0,1){20}}
\put(30,10){\circle*{3}}
\bezier{200}(30,-4)(32,-2)(30,0)
\bezier{200}(30,0)(28,2)(30,4)
\put(30,4){\line(0,1){2}}
\put(28.6,4){$\wedge$}
\put(40,10){\circle*{3}}
\bezier{200}(40,-4)(42,-2)(40,0)
\bezier{200}(40,0)(38,2)(40,4)
\put(40,4){\line(0,1){2}}
\put(38.6,4){$\wedge$}
\put(50,10){\circle*{1}}
\put(60,10){\circle*{1}}
\put(70,10){\circle*{1}}
\put(50,1){\circle*{1}}
\put(60,1){\circle*{1}}
\put(70,1){\circle*{1}}
\put(80,10){\circle*{3}}
\bezier{200}(80,-4)(82,-2)(80,0)
\bezier{200}(80,0)(78,2)(80,4)
\put(80,4){\line(0,1){2}}
\put(78.6,4){$\wedge$}
\bezier{200}(90,0)(100,10)(90,20)
\put(90,0){\line(0,1){20}}
\put(10,10){\dashbox(90,0)}
\put(99,9){$>$}
\end{picture}
\end{center}
\vspace{-20mm}
\begin{center}
{The general setting for a quantum computation based on Cavity QED : \\
the dotted line means a single photon inserted in the cavity and \\
all curves mean external laser fields subjected to atoms}
\end{center}
\vspace{5mm}
See \cite{FHKW1}, \cite{FHKW2} in detail. 
Usually it is based on two level system of atoms. However, 
to take a multi level system of them into consideration may be better from 
the view point of decoherence which is a severe problem in quantum 
computation.

It seems that the universal Yang--Mills theory developed in this paper 
gets on well with a quantum computation based on four level system, 
see for example \cite{KF4}, \cite{RZ}.

The gauge theory is useful even in quantum computation. This has been pointed 
out by Fujii \cite{KF1}, \cite{KF2} in the context of Holonomic Quantum 
Computation. 
In a forthcoming paper we want to point out a deeper relation between 
the gauge theory and some quantum computation (Cavity QED quantum computation 
if possible).



\begin{thebibliography}{99}
%
\bibitem{YM}C. N. Yang and R. L. Mills :
\newblock Conservation of isotopic spin and isotopic gauge invariance, 
\newblock Phys. Rev. 96 (1954), 191.
%
\bibitem{IM}M. Ikeda and Y. Miyachi :
\newblock On an extended framework for the description of elementary 
particles, 
\newblock Prog. Theor. Phys. 16 (1956), 537.
%
\bibitem{WY}T. T. Wu and C. N. Yang :
\newblock Concept of nonintegrable phase factors and global formulation  
of gauge fields, 
\newblock Phys. Rev. D 12 (1975), 3845.
%
\bibitem{BI}M. Born and L. Infeld :
\newblock Foundations of the new field theory, 
\newblock Proc. Royal Soc. (London), A 144 (1934), 425.
%
\bibitem{MN}M. Nakahara : 
\newblock Geometry, Topology and Physics, 
\newblock IOP Publishing Ltd, 1990.
%
\bibitem{BPST}A. A. Belavin, A. M. Polyakov, A. S. Schwartz and 
Yu. S. Tyupkin : 
\newblock Pseudoparticle solutions of the Yang--Mills equations, 
\newblock Phys. lett. B 59 (1975), 85.
%
\bibitem{AC}A. Connes : 
\newblock Noncommutative Geometry, 
\newblock Academic Press, 1994.
%
\bibitem{Nek}N. A. Nekrasov : 
\newblock Noncommutative Instantons Revisited, 
\newblock Commun. Math. Phys. 241 (2003) 143, 
\newblock hep-th/0010017.
%
\bibitem{FOS}K. Fujii, H. Oike and T. Suzuki :
\newblock in preparation.
%
\bibitem{KF3}K. Fujii : 
\newblock A Modern Introduction to Cardano and Ferrari Formulas 
in the Algebraic Equations, 
\newblock quant-ph/0311102. 
%
\bibitem{FHKW1}K. Fujii, K. Higashida, R. Kato and Y. Wada : 
\newblock Cavity QED and Quantum Computation in the Weak Coupling Regime, 
\newblock J. Opt. B: Quantum and Semiclass. Opt, 6(2004) 502, 
\newblock quant-ph/0407014. 
%
\bibitem{FHKW2}K. Fujii, K. Higashida, R. Kato and Y. Wada : 
\newblock Cavity QED and Quantum Computation in the Weak Coupling Regime II : 
Complete Construction of the Controlled--Controlled NOT Gate, 
\newblock to appear in the book "Quantum Computing : New Research", 2006, 
Nova Science Publishers, Inc (USA), 
\newblock quant-ph/0501046. 
%
\bibitem{KF4}K. Fujii : 
\newblock Study on Dynamics of N Level System of Atom by Laser Fields, 
\newblock quant-ph/0512126.
%
\bibitem{RZ}V. Ramakrishna and H. Zhou : 
\newblock On The Exponential of Matrices in $su(4)$, 
\newblock math-ph/0508018.
%
\bibitem{KF1} K. Fujii : 
\newblock Note on Coherent States and Adiabatic Connections, Curvatures, 
\newblock J. Math. Phys. 41(2000), 4406, 
\newblock quant-ph/9910069. 
%
\bibitem{KF2}K. Fujii : 
\newblock Introduction to Grassmann Manifolds and Quantum Computation, 
\newblock J. Applied Math, 2(2002), 371, 
\newblock quant-ph/0103011. 
%
\end{thebibliography}
\end{document}